\documentclass[reprint, floatfix,superscriptaddress, longbibliography,amsmath,amssymb,aps,prx,showkeys,
]{revtex4-2}
\makeatletter

\def\set@curr@file#1{%
 \begingroup
 \escapechar\m@ne
 \xdef\@curr@file{\expandafter\string\csname #1\endcsname}%
 \endgroup
}
\def\quote@name#1{"\quote@@name#1\@gobble""}
\def\quote@@name#1"{#1\quote@@name}
\def\unquote@name#1{\quote@@name#1\@gobble"}
\makeatother
\usepackage{graphicx}% Include figure files
\usepackage{dcolumn}% Align table columns on decimal point
\usepackage{bm}% bold math
\usepackage{xfrac}
\usepackage{siunitx}
\usepackage{amsmath}

\usepackage[utf8]{inputenc}
\usepackage[T1]{fontenc}
\usepackage[export]{adjustbox}
\usepackage[colorinlistoftodos,prependcaption,textsize=tiny]{todonotes}
\usepackage{gensymb}
\usepackage{lineno}
\usepackage{capt-of}
\usepackage{afterpage}
\usepackage{nonfloat}
\usepackage{filecontents}
\presetkeys%
 {todonotes}%
 {inline, backgroundcolor=yellow}{}
\usepackage{xr}
\makeatletter
\newcommand*{\addFileDependency}[1]{% argument=file name and extension
\typeout{(#1)}% latexmk will find this if $recorder=0
\@addtofilelist{#1}
\IfFileExists{#1}{}{\typeout{No file #1.}}
}\makeatother
\newcommand*{\myexternaldocument}[1]{%
\externaldocument{#1}%
\addFileDependency{#1.tex}%
\addFileDependency{#1.aux}%
}
\myexternaldocument{SM_EDITED}
\usepackage{color}
\newcounter{aqctr}
\newenvironment{author-query}
{\refstepcounter{aqctr}\par\vspace{\baselineskip}\noindent
\color{red}\textbf{Author Query/Comment AQ \arabic{aqctr}.}}
{\par\vspace{\baselineskip}\normalcolor}

\begin{document}
\preprint{APS/123-QED}
%TC:ignore
\title{Thermodynamic Bifurcations of Boiling in Solid-State Nanopores}% Force line breaks with \\

%\author[1]{Author A\thanks{A.A@university.edu}}
%\author[1]{Author B\thanks{B.B@university.edu}}
%\author[1]{Author C\thanks{C.C@university.edu}}
%\author[2]{Author D\thanks{D.D@university.edu}}
%\author[2]{Author E\thanks{E.E@university.edu}}
%\affil[1]{Department of Computer Science, \LaTeX\ University}
%\affil[2]{Department of Mechanical Engineering, \LaTeX\ University}
%\author[1]{Alice Smith}
%\author[2]{Bob Jones}
%\affil[1]{Department of Mathematics, University X}
%\affil[2]{Department of Biology, University Y}
\author{Soumyadeep Paul}
\email{Corresponding author: soumyadeep.paul@thml.t.u-tokyo.ac.jp}
\affiliation{%
Department of Mechanical Engineering, The University of Tokyo, 7-3-1, Hongo, Bunkyo-ku, Tokyo 113-8656, Japan
}%
\author{Yusuke Ito}%
\affiliation{%
Department of Mechanical Engineering, The University of Tokyo, 7-3-1, Hongo, Bunkyo-ku, Tokyo 113-8656, Japan
}%
\author{Wei-Lun Hsu}%
\affiliation{%
Department of Mechanical Engineering, The University of Tokyo, 7-3-1, Hongo, Bunkyo-ku, Tokyo 113-8656, Japan
}%
\author{Hirofumi Daiguji}%
\email{Corresponding author: daiguji@thml.t.u-tokyo.ac.jp}
\affiliation{%
Department of Mechanical Engineering, The University of Tokyo, 7-3-1, Hongo, Bunkyo-ku, Tokyo 113-8656, Japan
}%

%\affiliation{%
%Department of Mechanical Engineering, The University of Tokyo\\ 7-3-1, Hongo, Bunkyo-ku, Tokyo 113-8656, Japan
%}%


%\affiliation{
%\affiliation{

 % but any date may be explicitly specified

\begin{abstract}
%TC:ignore
Boiling heat transfer is the basis of many commonly used cooling techniques. In cooling of electronic devices, for example, it is desirable to further miniaturize heat exchangers to achieve higher heat transfer, and thus it is necessary to understand boiling phenomena on shorter spatial and temporal scales. This is especially challenging at the nanometer scale because conventional imaging techniques cannot capture the dynamics of nanobubbles, owing to the Abbe diffraction limit. Here in this research, we utilize the nanopore Joule heating system that enables the generation of nanobubbles and simultaneous diagnosis of their nanosecond resolution dynamics using resistive pulse sensing. When a bias voltage is applied across a silicon nitride nanopore immersed in an aqueous salt solution, Joule heat is generated owing to the flow of ionic current. With increasing voltage, the Joule heating intensifies, and the temperature and entropy production in the pore increase. Our sensing results show that nanopore boiling follows the theory of minimum entropy production and attempts to settle to a minimum dissipative state.
This results in two boiling bifurcations corresponding to the transition between different boiling states. These characteristics of nanopore boiling are represented by an ``M''-shaped boiling curve, experimentally obtained from the Joule heat variation with the applied voltage. A theoretical framework is proposed to model the thermodynamics of nanopore bubbles and estimate the system dissipation which explains the four arms of the ``M''-shaped boiling curve. The present study reveals that the utilization of nanopore boiling as a benchmark platform offers a valuable means for investigating the intricate boiling phenomenon and its correlation with nanoscale bubble dynamics. This would provide much-needed fundamental insights into the chaotic transition boiling regime, which is least understood.

% Boiling heat transfer on wetted hydrophilic surfaces plays a crucial role in microchannel heat sinks for electronic cooling and metal heat treatment industries. Especially at the nanoscale, traditional imaging techniques fail to capture the nanosecond resolution dynamics due to the Abbe diffraction limit. To fill this knowledge gap, we utilize nanopore technology and resistive pulse sensing for boiling diagnostics at megahertz bandwidth. When bias voltages are applied across a silicon nitride nanopore immersed in aqueous salt solutions, the flow of ionic current results in Joule heat generation. With increasing voltage, the Joule heating intensifies, increasing the temperature and entropy production within the pore. Following the theory of minimum entropy production, the nanopore system tries to settle down to the state of least dissipation, through fluctuation-driven activation energy processes like bubble nucleation. This leads to nanopore boiling bifurcations along with transition regimes and catastrophes which are summarized in an experimentally obtained 'M'-shaped boiling curve and explained through variations in a theoretically estimated system dissipation function. 

%\begin{description}
%\item[Keywords]
%You may use the \texttt{description} environment to structure your abstract
%\end{description}
\end{abstract}
%TC:endignore
\maketitle
%\tableofcontents

\section{INTRODUCTION}\label{sec:level1}
Boiling is an unsteady nonequilibrium process that plays a vital role in diverse engineering applications,  including thermal desalination~\cite{liu2022heat}, power generation~\cite{Chen2009},  cooling of electronic devices~\cite{Dhillon2015, paul2021analysis, paul2020single, thome2007bubble}, inkjet printing~\cite{Asai1989}, and spray quenching~\cite{jiang2022inhibiting}. Much attention has been paid to extending the nucleate boiling regime during which a huge amount of heat flux can be dissipated from a solid surface through microlayer evaporation~\cite{dhir1998boiling, zhang2023unifying, burevs2021modelling, Thome2004}. However, there is a lack of comprehensive understanding of boiling at the nano- to microscale, where much of the process takes place below the optical limit (Abbe diffraction limit)~\cite{lavino2021surface, paul2022boiling}.

In recent years, both pool boiling experiments~\cite{popov2018detection} and nanoparticle boiling~\cite{Lombard2016, maheshwari2018dynamics, holyst2017evaporation} studies have shown that vapor nanobubbles of dimensions below the roughness scale can exist stably on a hydrophilic boiling surface. Several plausible and self-consistent theoretical explanations, such as contact-line pinning~\cite{lohse2015surface}, Knudsen effects~\cite{seddon2011knudsen}, and dynamic equilibrium~\cite{Brenner2008}, have been proposed to explain the stability of nanobubbles created by dissolved gas. However, their applicability to vapor bubble stability in the context of boiling heat transfer is still not known. Vapor bubble dynamics, which dominate boiling heat transfer, differ significantly from gas bubble dynamics, particularly in the mechanisms of interfacial heat and mass transport and the associated time scales~\cite{Prosperetti2017}. Hence, to handle the complexity associated with the nonequilibrium thermodynamics and nonlinear dynamics~\cite{Nicolis1989-NICECA} of nanoscale boiling, we need to start from a thermodynamic stability analysis of vapor nanobubbles. In the case of macroscale pool boiling, Shoji and co-workers~\cite{shoji2004studies, chai2001boiling, chai2002self} applied Prigogine's theory of self organization~\cite{nicolis1977self} to explain the ``S''-shaped boiling curve and clarified the entropic origins of the two bifurcation points: the onset of nucleate boiling and the onset of stable film boiling. In a similar vein, we study the thermodynamic bifurcations of nanoscale boiling, based on well-controlled bubble dynamics experiments using a nanopore Joule heating platform~\cite{Levine2016, Nagashima2014, paul2020single, paul2021analysis, paul2022boiling}, which allows us to capture vapor nanobubble dynamics at nanosecond resolutions.

\begin{figure*}[!t]
%\begin{p}
%\centering
\includegraphics[width=1\textwidth,keepaspectratio,angle=0]{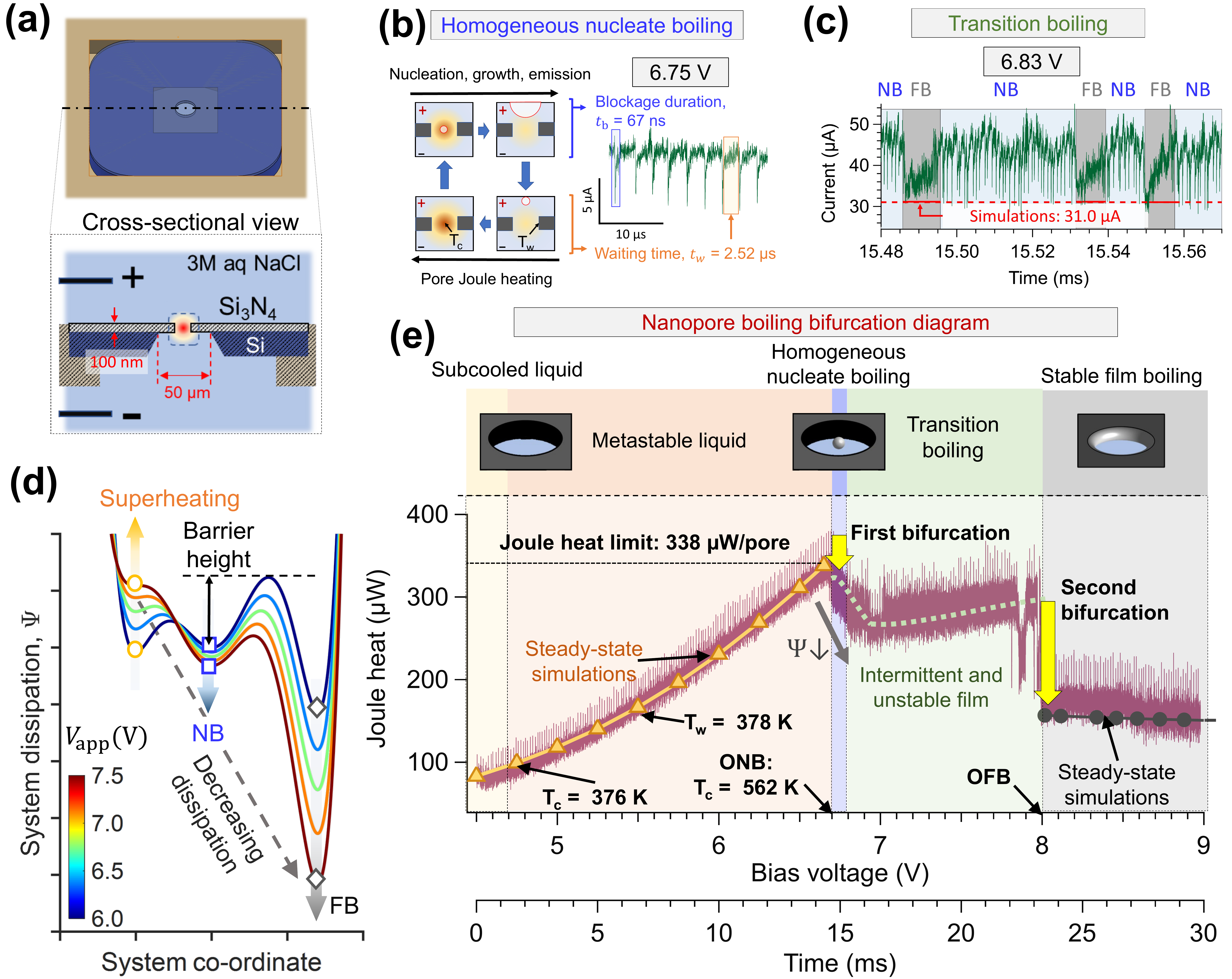}
\caption{\label{fig:1} Boiling bifurcations in a 199~nm diameter nanopore. (a) Schematic of nanopore Joule heating setup. Boiling signals are captured simultaneously using a passive (more noise) and an active (less noise, owing to higher capacitance) oscilloscope probe. (b) Homogeneous nucleate boiling comprises discrete bubble events, resulting in periodic bubble blockage signals. The plot shows filtered current from the active probe. $T_\mathrm{c}$ and $T_\mathrm{w}$ denote the pore center  and  pore surface temperatures, respectively. (c) Transition boiling shows intermittent film bubbles disrupting nucleate boiling. The plot shows filtered current from the passive probe. (d) A phenomenological plot of the system dissipation function versus system coordinates (superheating, nucleate, and film boiling) explains the first bifurcation point. (e) An ``M''-shaped nanopore boiling curve showing the distinct zones of subcooled liquid, metastable liquid, nucleate boiling, transition boiling, and stable film boiling. The two bifurcation points at the onset of nucleate boiling (ONB) and the onset of stable film boiling (OFB) are indicated.}
%\end{p}
\end{figure*}
Using this platform, we have previously characterized nucleate, transition, and film boiling regimes based on acoustic and resistive pulse sensing signals~\cite{paul2022boiling}. In the present study, we construct an ``M''-shaped boiling curve [Fig.~\ref{fig:1}(e)] summarizing the boiling regimes for a 199~nm diameter solid-state nanopore (Fig.~\ref{fig:a1} in the Supplemental Material~\cite{supp}), explaining the two bifurcation points based on the variation of system dissipation. Furthermore, we shed light on the thermodynamics of the film bubble interface, elucidating the role of (i) Knudsen bubble stability and (ii) its out-of-equilibrium behavior on the system dissipation and boiling curve for the nanopore system.

The nanopore was fabricated using focused ion beam etching on a 100~nm thick suspended silicon nitride membrane chip [Fig.~\ref{fig:1}(a)]. When bias voltages are applied across the nanopore after its immersion in 3M NaCl solution, the flow of ionic current $I$ results in focused Joule heating in the nanopore liquid [Fig.~\ref{fig:1}(a)], leading  to bubble nucleation~\cite{Nagashima2014, Levine2016, paul2020single}. The formation of bubbles within the nanopore hinders ionic current flow, which is captured in a high-bandwidth oscilloscope [Figs.~\ref{fig:1}(b) and~\ref{fig:1}(c)]. For detailed information about experimental methods, we refer to our earlier works~\cite{paul2020single, paul2022boiling}. Discrete and periodic current blockage signals represent the nucleation of a homogeneous vapor bubble  at the pore center, followed by its thermal growth and ejection, and finally its shrinkage at the pore entrance [Fig.~\ref{fig:1}(b)]~\cite{paul2022boiling}, following which there is a waiting/heating period before the subsequent bubble nucleation. We term this sequential bubble generation as nanopore nucleate boiling (NB). As the bias voltage is intensified, the nanopore surface becomes blanketed by a torus-shaped vapor film (film boiling, FB), causing a continuous decrease in baseline current~\cite{paul2022boiling}. In Fig.~\ref{fig:1}(c), we find that when the voltage is increased to 6.83~V, a step-like decrease in baseline current (corresponding to FB) occurs multiple times during periodic current dips (corresponding to NB). This resembles intermittent film boiling, which falls under transition boiling. Following the theory of minimum entropy production, the nanopore system attempts to settle to a minimum dissipative state (superheating, NB, or FB) through fluctuation-driven activation processes such as bubble nucleation and vapor film formation. The tristable dissipation well diagram in Fig.~\ref{fig:1}(d) shows increasing dissipation wells of superheating and decreasing dissipation wells of NB and FB with bias voltage. This explains the gradual system transition to nucleate and then intermittent film boiling as observed in experiments.
 
The variation of Joule heat dissipation with bias voltage is summarized in the form of an ``M''-shaped boiling curve [Fig.~\ref{fig:1}(e)]. In the following sections, after an explanation of the two bifurcations [ONB and OFB in Fig.~\ref{fig:1}(e)] in the ``M''-shaped boiling curve, the superheating and stable film boiling regimes are explained through steady-state continuum simulations of Joule heating, using kinetic theory boundary conditions for interfacial heat and mass transfer (Fig.~\ref{fig:2}). Furthermore, nucleate boiling and transition boiling, which constitute the unsteady boiling regimes, are described. The dissipation function, which represents the entropy generation of the nanopore system, has been theoretically calculated for all the boiling regimes and shown to be ``M''-shaped with respect to bias voltage.

\section{Results and Discussion}\label{secIII}
\subsection{Boiling bifurcation diagram}\label{secIIIA}
In the present study, we apply a ramp voltage pulse [linearly increasing from 4.5~V to 9~V across 30~ms, $V_\mathrm{app}= 4.5 + 150t~[\mathrm{s}]$;  Fig.~\ref{fig:a2}(a) in the Supplemental Material~\cite{supp}] across the nanopore, which was initially in thermal equilibrium at an ambient temperature of 298.15~K. The variation of mean Joule heat generation inside the nanopore during the 30~ms and the boiling modes at discrete time points are studied to explain the global and local aspects of nanoscale boiling. Owing to the very slowly changing bias voltage (nonequilibrium boundary condition), each boiling mode reached during the ramp voltage pulse can be assumed to be in a quasi-steady state. For reference, the thermal relaxation time for the 199~nm nanopore [$\tau_\mathrm{th}\sim R_\mathrm{p}^2/D_\mathrm{th}=\mathcal{O}$(100~ns)] and the lifetime of nucleate bubbles [$t_\mathrm{b}=67$~ns, Fig.~\ref{fig:1}(b)] is much shorter than the millisecond-order voltage pulse. Here, $R_\mathrm{p}$ is the pore radius and $D_\mathrm{th}$ is the thermal diffusivity of water. 

Similar to pool boiling, nanopore boiling is a nonequilibrium phase change process, but the critical difference is the boundary condition that causes nonequilibrium~\cite{prigogine1978time}. In pool boiling experiments~\cite{dhir1998boiling}, the externally controlled wall temperature is the boundary condition that makes the system nonequilibrium~\cite{chai2002self}, whereas in nanopore boiling, the bias voltage acts as the nonequilibrium boundary condition. Accordingly, we plot the nanopore boiling curve with the applied bias voltage $V_\mathrm{app}$ on the horizontal axis and the resulting Joule heat dissipation $V_\mathrm{app} \times I$ on the vertical axis [Fig.~\ref{fig:1}(e)]. It should be noted here that the creation of Joule heat in a conductor is a significant source of entropy production~\cite{onsager1931reciprocal, demirel2007nonequilibrium} and accordingly an appropriate quantitative measure of the nonequilibrium nature of the nanopore system. Being an open dissipative system, nanopore boiling tries to self-organize and minimize entropy production from Joule heat dissipation. This leads to the two bifurcation points: onset of homogeneous nucleate boiling and onset of stable nanotorus film boiling [Fig.~\ref{fig:1}(e)]. At both these points, the slope of Joule heat versus voltage changes from positive to negative, i.e., a decreasing trend of dissipation is achieved, which is a more favorable thermodynamic pathway for the system to progress.

%The first bifurcation can be conceptually understood through the tristable potential well schematic shown in Fig.~\ref{fig:1}d. At low voltages, when the pore is fully wetted, the well for superheating rises with increasing voltages, while the nucleate boiling (NB) well dips, making NB a more  dissipative structure. Also, the barrier height for the system transition from superheating to nucleate boiling decreases. Next, with increasing voltages, the periodicity of bubble nucleation increases which decreases the NB well height. However, there is an upper limit on the nucleation frequency as the nanopore hotspot only allows one bubble to nucleate at a time~\cite{Nagashima2014, paul2020single}. Also at this point, nucleation of heterogeneous bubbles on the pore surface, coalescing to a vapor film becomes easier as the wall temperature, $T_\mathrm{w}$ also rises with bias voltages (Fig.~\ref{fig:a2}c). At the same time, the film bubble occludes more of the pore volume, which allows the system to have a much lower dissipation than NB. This results in the gradual transition from nucleate towards film boiling.\par

% \subsection{Theoretical model}\label{secIIIBmid}
\subsection{Steady states}\label{secIIIB}
\subsubsection{Superheating}\label{secIIIBi}
As the voltage is increased from 4.5~V to 6.65~V,  the nanopore current increases continuously [Fig.~\ref{fig:a2}(a) in the Supplemental Material~\cite{supp}]. As a result, the generation of Joule heat within the nanopore increases monotonically and constitutes the left arm of the ``M''-shaped boiling curve. We use continuum simulations to implement a Joule heating model, which has been described in our earlier papers~\cite{paul2020single, paul2022boiling} and is restated in Sec.~S2 in the Supplemental Material~\cite{supp}. Using the Joule heating model, the steady-state nanopore temperatures at the pore center ($T_\mathrm{c}$) and pore walls ($T_\mathrm{w}$) during superheating [Fig.~\ref{fig:a2}(c) in the Supplemental Material~\cite{supp}] are obtained. The contour plots in Figs.~\ref{fig:3}(a) and~\ref{fig:3}(b) show the spatial distribution of nanopore temperature for bias voltages of 6.0~V and 6.65 V, respectively. The theoretically obtained nanopore current and Joule heat generation [yellow lines with markers in Figs.~\ref{fig:a2}(a) and~\ref{fig:1}(e), respectively] fit well with the experimental results. According to the simulations, we find that $T_\mathrm{c}$ and $T_\mathrm{w}$ exceed the saturation temperature of 373.15~K at voltages lower than 5.5~V. However, no bubble signals are seen until 6.7125~V,  when $T_\mathrm{c}=562$ K is reached. This value is close to the kinetic limit for homogeneous nucleation, 575~K~\cite{Avedisian1985}. The enhanced range of metastability of the nanopores and the prevention of heterogeneous nucleation at low voltages were shown by Golovchenko and co-workers~\cite{Nagashima2014, Levine2016}, and were later explained by Paul \emph{et al.}~\cite{paul2020single} through a nucleation theory model based on ripening competition between homogeneous and heterogeneous bubble clusters. This unique characteristic is due to a large temperature difference between $T_\mathrm{c}$ and $T_\mathrm{w}$, which originates from the focused nature of Joule heating.

We also calculate the total system dissipation during superheating, $\Psi=\int_{V}{\psi}\,dV$, where the local dissipation in the liquid volume can be written as $\psi=Td_is/dt=T\dot{s}_\mathrm{gen}$~\cite{demirel2007nonequilibrium}. Here, $T$ is the liquid temperature and $d_is/dt$ is the entropy generation per unit volume, which can be expressed as
\begin{equation}
\frac{d_is}{dt}=\mathbf{J_\mathrm{u}}\cdot\nabla\left(\frac{1}{T}\right)+\frac{1}{T}\mathbf{J}\cdot\mathbf{E}.
\label{eq:1}
\end{equation}%NOTE FROM EDITOR: Please note the change from 10^{11}{} etc. to SI{e11} etc.  below: the latter gives the correct spacing between the number and the unit. Note also that \,\mathord{:}\, gives better spacing around the ``double dot product'' than a colon alone.
Here, $\mathbf{J_\mathrm{u}}=k_\mathrm{w}\nabla T$ is the heat flux in the liquid and $\mathbf{J}=\sigma\mathbf{E}$ is the ionic current, where $k_\mathrm{w}$ is the thermal conductivity and $\sigma$ is the electrical conductivity, and $\mathbf{E}=-\nabla\phi$ is the electric field generated in the nanopore liquid due to the applied bias voltage $\Delta\phi=V_\mathrm{app}$. The first and second terms on the right-hand side of Eq.~\eqref{eq:1} represent the entropy generation due to heat conduction and Joule heating, respectively. Here, we neglect the viscous dissipation term $\bm{\tau}\,\mathord{:}\, (\nabla \mathbf{u})\approx\mu({\partial u}/{\partial x})^2\approx\mu({u}/{R_\mathrm{p}})^2=\mathcal{O}(\SI{e11}{\watt\per\meter^3})$, which is much smaller than the dissipation originating from Joule heating, $\sigma{\vert\mathbf{E}\vert}^2=\mathcal{O}(\SI{e16}{\watt\per\meter^3})$ [see Fig.~\ref{fig:a35}(a) in the Supplemental Material~\cite{supp}], and heat conduction, $k_\mathrm{w}{\nabla T}^2/T=\mathcal{O}(\SI{e15}{\watt\per\meter^3})$ [see Fig.~\ref{fig:a35}(b) in the Supplemental Material~\cite{supp}]. Here, $u=\mathcal{O}(\SI{1}{\meter\per\second})$ is assumed, which is consistent with the nanopore electroosmotic velocity~\cite{di2022geometrically, wang2020joule}. Using Eq.~\eqref{eq:1}, we calculate the system dissipation $\Psi$, which increases monotonically with voltage during superheating [Fig.~\ref{fig:3}(g)]. The breakdown of dissipation sources is shown in Fig.~\ref{fig:a2}(c) in the Supplemental Material~\cite{supp}, where Joule heating in the liquid is the dominant source of dissipation, while heat conduction in the silicon nitride membrane makes the least contribution. The superheating regime is followed by periodic nucleate boiling and intermittent film boiling, during which the time-averaged dissipation decreases with voltage (discussed further in the following sections), causing the formation of a stable dissipative structure.

\begin{figure*}[!t]
\includegraphics[width=0.67\textwidth,keepaspectratio,angle=0]{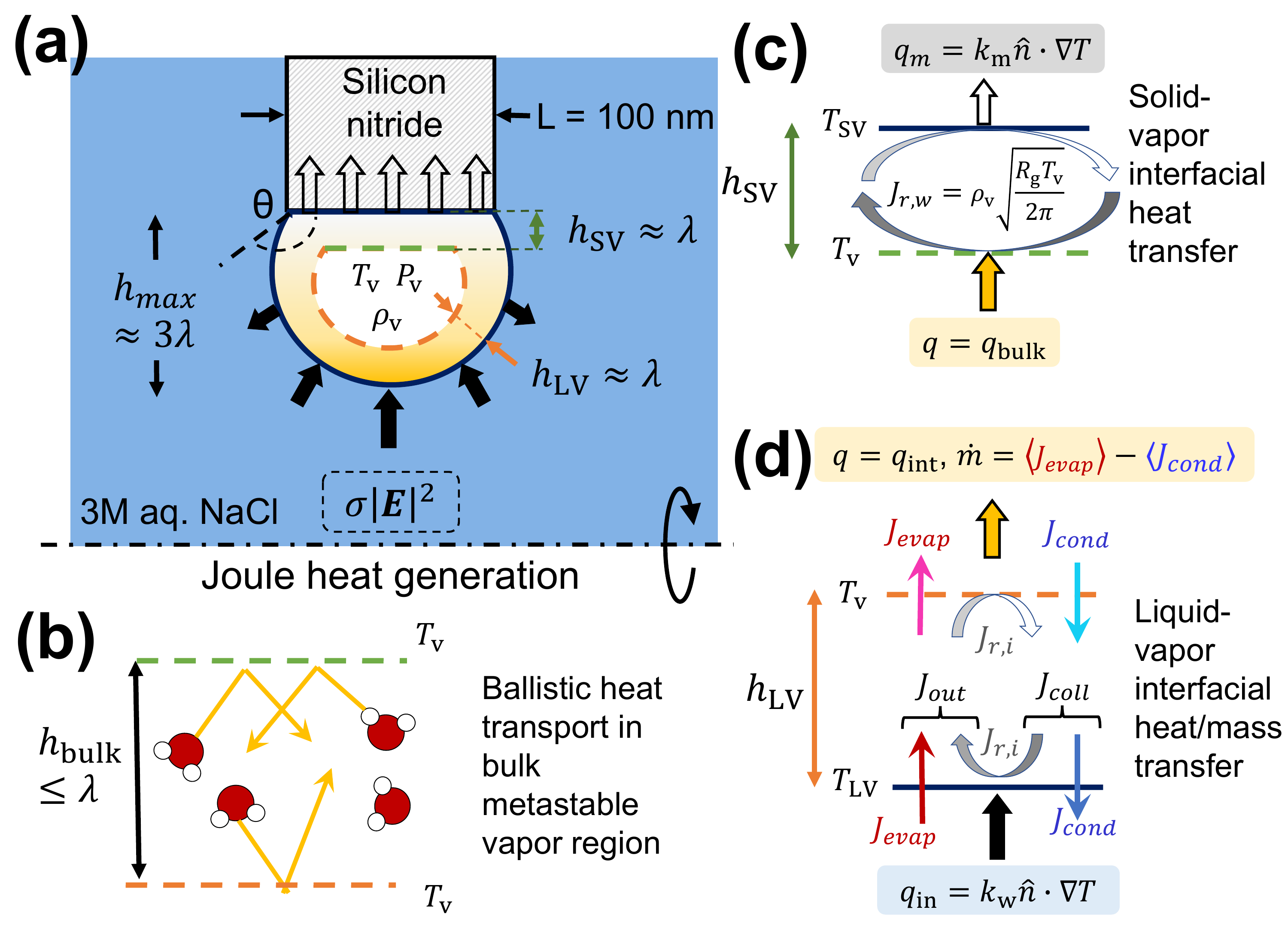}
\caption{\label{fig:2} Schematic of the model of heat transfer in the Knudsen vapor film bubble. (a) Nanopore torus bubble consisting of the liquid--vapor interface, the solid--vapor interface, and the bulk--vapor region between these two interfaces. (b) Ballistic heat flux within the bulk vapor region. (c) Explanation of  heat transfer at the solid--vapor interface, where heat transfer takes place through molecular reflection. (d) Heat transfer at the liquid--vapor interface, with evaporation, condensation, and reflection fluxes.}
%\end{p}
\end{figure*}

\subsubsection{Stable film boiling}\label{secIIIBii}
Following the second bifurcation at 8~V,   stable film boiling (SFB) commences. During this regime, a torus-shaped vapor film blankets the cylindrical pore surface and exists stably despite the steep temperature fields and complex heat flow within the nanopore. As the bubble occludes the pore volume, there are catastrophic decreases in the nanopore current [Fig.~\ref{fig:a2}(a) in the Supplemental Material~\cite{supp}] and Joule heat generation [Fig.~\ref{fig:1}(e)]. In our previous work~\cite{paul2022boiling}, we found that the magnitude of the current drop was proportional to the volume of the pinned torus bubble. This relationship was found to be valid for 340~nm, 420~nm, and 460~nm diameter pores. These trends strongly suggested the presence of a pinned torus bubble on the pore surface. In Fig.~\ref{fig:a2}(a) in the Supplemental Material~\cite{supp}, we find that as the voltage is increased, the nanopore current decreases, which can be explained by the bulging out of the vapor film pinned at the nanopore wall. In other words, the contact angle of the bubble [$\theta$ in Fig.~\ref{fig:2}(a)] increases with voltage. As the bubble occupies a larger fraction of the pore volume, both experimental [purple trace in Fig.~\ref{fig:1}(e)] and simulation [black line in Fig.~\ref{fig:1}(e)] studies of vapor film bulging out show that Joule heat dissipation is reduced. We study the exceptional stability of the torus bubble through a continuum model of heat and mass transfer and explain the bulging-out effect of the vapor film with increasing bias voltage. Details of the simulation results for stable film bubbles are given in Sec.~S3 in the Supplemental Material~\cite{supp}.

\begin{figure*}[!t]
%\begin{p}
%\centering
\includegraphics[width=1\textwidth,keepaspectratio,angle=0]{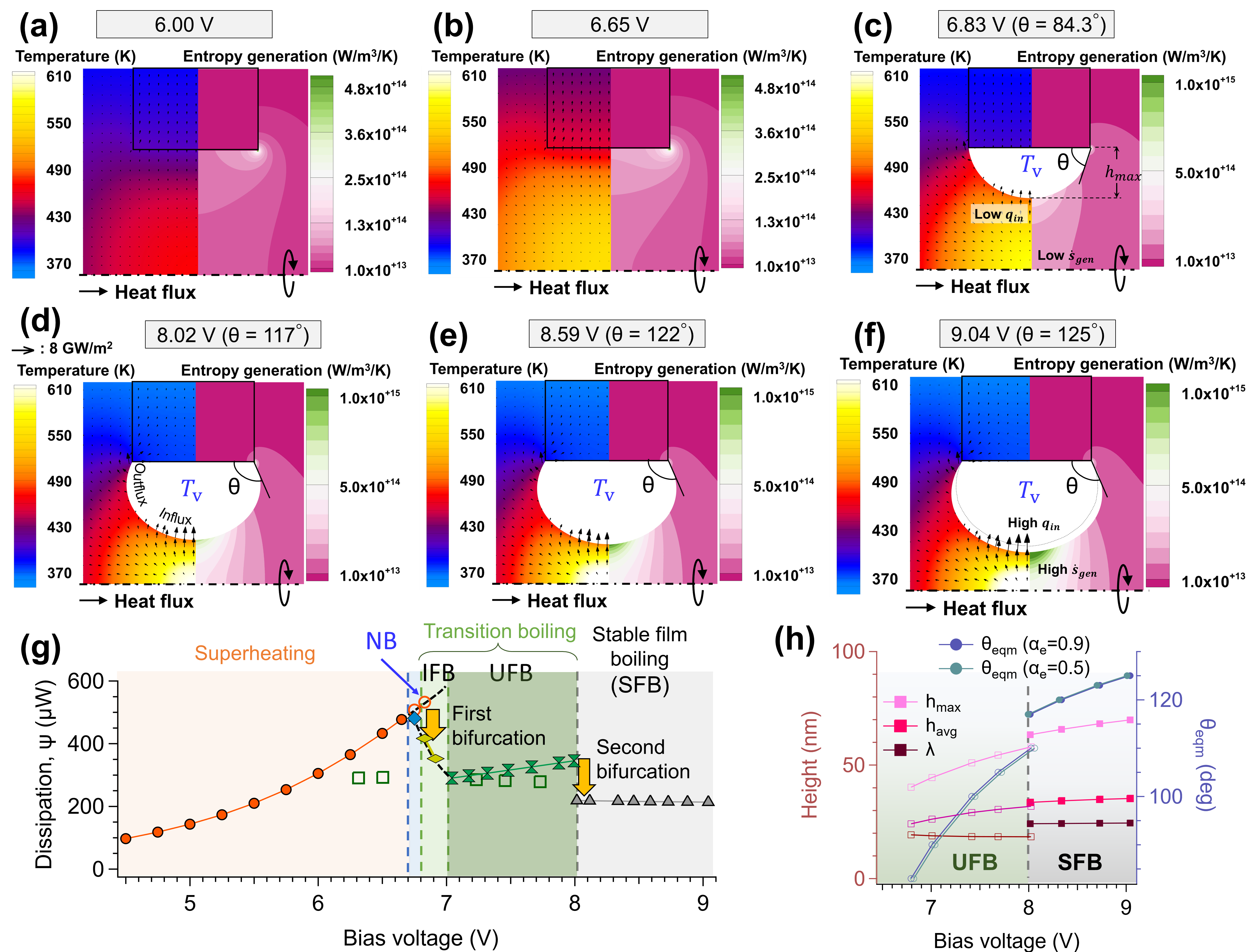}
\caption{\label{fig:3} Contour plots showing distributions of temperature (left) and entropy generation inside a nanopore during superheating at (a) 6.00~V and  (b) 6.65~V,  intermittent film boiling at (c) 6.83~V,  and  steady-state film boiling at (d) 8.02~V,  (e) 8.59~V, and (f) 9.04 V. The dotted line in (f) shows the smaller bubble shape at 8.02~V. The vectors in each subplot denote the heat flux. (g) System dissipation function calculated through continuum simulations. (h) Variations in height, contact angle, and mean free path of vapor molecules within the equilibrium film bubble for changing bias voltage.}
%\end{p}
\end{figure*}

Nonuniform and concentrated Joule heat generation [Fig.~\ref{fig:a35}(c) in the Supplemental Material~\cite{supp}] causes nonuniform thermal gradients on the torus bubble surface. The shape of the torus bubble is determined under the assumption that it is pinned to the surface of a cylindrical pore of  thickness  $L=100$~nm with uniform mean curvature $K$ [Fig.~\ref{fig:2}(a)]. A further description can be found in our previous paper~\cite{paul2022boiling}. Under these constraints, the contact angle $\theta$ is the only free geometric parameter that controls the bubble shape. The larger the angle $\theta$, the more the bubble expands [Figs.~\ref{fig:3}(d)--\ref{fig:3}(f)], blocking much of the pore volume and limiting current flow and Joule heat generation.

Now, the tip-to-base height of the bubble satisfies $h_\mathrm{max}\approx3\lambda$, where $\lambda$ is the thermal mean free path of vapor molecules, which satisfies $\lambda=1.922\lambda_\mathrm{HS}=1.922 m/(\sqrt{2}\,\pi\rho_\mathrm{v}d^2)$~\cite{dongari2011modeling}. $\lambda_\mathrm{HS}$ is the mean free path per hard sphere model. Here, $m$ is the mass of one vapor molecule, $\rho_\mathrm{v}$ is the vapor density, and $d$ is the collision diameter, assumed to be $2.7$~\AA. For the Knudsen bubble, the volume can be divided into liquid--vapor (LV) and solid--vapor (SV) kinetic interfaces and the vapor bulk regions. Each interface has a height of one mean free path ($h_\mathrm{LV}\sim h_\mathrm{SV}\sim\lambda$), and the vapor bulk region has a length of less than one mean free path [$h_\mathrm{bulk}<\lambda$ in Fig.~\ref{fig:2}(b)]. Figure~\ref{fig:3}(h) shows the height variations of the equilibrium film bubble at varying voltages, where we find the average bubble height of the SFB to be $h_\mathrm{avg}\approx1.16\lambda$. Hence, we consider temperature drops across the LV and SV interfaces (net thickness  $2\lambda$) to account for the net temperature drop across the bubble.

Considering the heat transfer at the SV interface to be governed by molecular reflections [Fig.~\ref{fig:2}(c)], we express the heat flux as
\begin{equation}
q_\mathrm{m}=-k_\mathrm{m}\frac{\partial T}{\partial n}=q_\mathrm{bulk}=\rho_\mathrm{v}\sqrt{\frac{R_\mathrm{g} T_\mathrm{v}}{2\pi}}\,c_\mathrm{v}(T_\mathrm{v} - T_\mathrm{SV}),
\label{eq:12}
\end{equation}
where $R_\mathrm{g}$ is the universal gas constant, and $k_\mathrm{m}$ is the thermal conductivity of the silicon nitride membrane, taken as \SI{3.2}{\watt\per\meter\per\kelvin}~\cite{Levine2016}. $q_\mathrm{bulk}$ is the incoming heat flux from the vapor bulk region to the SV interface. $\rho_\mathrm{v}$, $c_\mathrm{v}$, and $U_\mathrm{v}$ are the vapor density, specific heat at constant volume, and internal energy obtained from the phase diagram of water according to the IAWPS formulation~\cite{Wagner2002} [Fig.~\ref{fig:a33}]. As shown in Fig.~\ref{fig:2}(d), the heat flux balance at the LV interface can be written as
\begin{align}
q_\mathrm{in}={}& q_\mathrm{evap} -  q_\mathrm{cond} + q_\mathrm{ref} + h_\mathrm{fg}\dot{m} \nonumber \\
={}&  \alpha_\mathrm{e}\rho_\mathrm{v,sat}U_\mathrm{v,sat}\sqrt{\frac{R_\mathrm{g} T_\mathrm{LV}}{2\pi}}-\alpha_\mathrm{c}\rho_\mathrm{v}U_\mathrm{v}\sqrt{\frac{R_\mathrm{g} T_\mathrm{v}}{2\pi}} \nonumber \\
&+(1-\alpha_\mathrm{e})\rho_\mathrm{v,sat}\sqrt{\frac{R_\mathrm{g} T_\mathrm{LV}}{2\pi}}\,c_\mathrm{v}(T_\mathrm{LV} - T_\mathrm{v})+ h_\mathrm{fg}\dot{m}.
\label{eq:2}
\end{align}
Here, the subscript ``sat'' denotes the vapor phase evaporating at the liquid side of the LV interface, which is assumed to be saturated at $T_\mathrm{LV}$. 
% Owing to the high value of $k_\mathrm{m}$ compared with the thermal conductivities of water and vapor, we assume that $T_\mathrm{SV}$ is uniform on the pore surface. 
The latent heat of evaporation or condensation ($h_\mathrm{fg}\dot{m}$) associated with the net emission or trapping of molecules at the liquid side of the LV interface~\cite{Cao2011} consumes a significant amount of the incoming heat flux from the bulk liquid [$q_\mathrm{in}=k_\mathrm{w}\hat{n}\cdot\nabla T$ in Figs.~\ref{fig:2}(d) and~\ref{fig:3}(f)]. The remaining heat ($q_\mathrm{int}=q_\mathrm{evap} -  q_\mathrm{cond} + q_\mathrm{ref}$) is transferred through the molecular fluxes of evaporation, condensation, and reflection at the LV interface  [$J_\mathrm{evap}$, $J_\mathrm{cond}$, and $J_\mathrm{r,i}$ as shown in Fig.~\ref{fig:2}(d)]. The molecular velocities in these fluxes are defined based on Maxwellian distributions~\cite{aursand2019comparison, Kobayashi2018, Kobayashi2016}. Accordingly, $\langle J_\mathrm{out}\rangle=\rho_\mathrm{v,sat}\sqrt{{R_\mathrm{g} T_\mathrm{LV}}/{2\pi}}$ and $\langle J_\mathrm{coll}\rangle=\rho_\mathrm{v}\sqrt{{R_\mathrm{g} T_\mathrm{v}}/{2\pi}}$, alongwith $\langle J_\mathrm{evap}\rangle=\alpha_\mathrm{e}\langle J_\mathrm{out}\rangle$ and $\langle J_\mathrm{cond}\rangle=\alpha_\mathrm{c}\langle J_\mathrm{coll}\rangle$~\cite{Cao2011}.
% A more detailed description of the equations can be found in the Supplemental Material~\cite{supp}. 
The net evaporation/condensation mass flux across the LV interface can then be written as
\begin{equation}
\dot{m} = \alpha_\mathrm{e}\rho_\mathrm{v,sat}\sqrt{\frac{R_\mathrm{g} T_\mathrm{LV}}{2\pi}}-\alpha_\mathrm{c}\rho_\mathrm{v}\sqrt{\frac{R_\mathrm{g} T_\mathrm{v}}{2\pi}}\,,
\label{eq:13}
\end{equation}
where $\alpha_\mathrm{e}$ and $\alpha_\mathrm{c}$ are the evaporation and condensation accommodation coefficients, respectively. For the results shown in Fig.~\ref{fig:3}, $\alpha_\mathrm{e}=0.9$ is uniformly applied on the LV interface. Applying the local balance of fluxes on the LV interface ($\langle J_\mathrm{out}\rangle-\langle J_\mathrm{coll}\rangle=\langle J_\mathrm{evap}\rangle-\langle J_\mathrm{cond}\rangle$), $\alpha_\mathrm{c}$ can be written as
\begin{equation}
\alpha_\mathrm{c} = 1-\frac{(1-\alpha_\mathrm{e})\rho_\mathrm{v,sat}}{\rho_\mathrm{v}}\sqrt{\frac{T_\mathrm{LV}}{T_\mathrm{v}}}\,.
\label{eq:14}
\end{equation}
We also apply the Laplace equation on the LV surface along with  closure conditions of no heat and mass accumulation to solve for the steady-state bubble size and temperature:
\begin{equation}
\begin{aligned}
P_\mathrm{v}&=K\gamma+P_\mathrm{w},  \\
\iint_{S_\mathrm{LV}} \dot{m} dS &= 0,   \\
\iint_{S_\mathrm{LV}} q_\mathrm{in}\, dS &= \iint_{S_\mathrm{SV}} q_\mathrm{m} \,dS.
\end{aligned}
\label{eq:4}
\end{equation}
Here, $P_\mathrm{v}$ and $P_\mathrm{w}$ are the vapor and liquid pressures, respectively. $S_\mathrm{LV}$ and $S_\mathrm{SV}$ are the liquid--vapor and solid-vapor interfaces respectively as shown in Fig.~\ref{fig:a22} (inset) in SI~\cite{supp}. $\gamma$ is the surface tension of water defined at $T_\mathrm{v}$. We solve the governing equations for Joule heating of nanopores described in Sec.~S2 in the Supplemental Material~\cite{supp} under the above-described heat and mass flux boundary conditions on a finite volume mesh [Figs.~\ref{fig:a22}] using the multiphysics software,~\textit{arb}~\cite{Harvie2012}. This yields the temperature distribution of nanopores for varying bias voltage [Figs.~\ref{fig:3}(d)--\ref{fig:3}(f)]. We find that the film bubble expands [$\theta_\mathrm{eqm}$ increases from $117\degree$ to $125\degree$ as shown in Figs.~\ref{fig:3}d--\ref{fig:3}(f) and~\ref{fig:3}(h)] to satisfy the closure equations~\eqref{eq:4} for increasing voltage. The right panels in Figs.~\ref{fig:3}(d)--\ref{fig:3}(f) show the net entropy generation in the bulk liquid and solid regions, which is a combination of Joule heating [Fig.~\ref{fig:a35}(c) in the Supplemental Material~\cite{supp}] and thermal gradient [Fig.~\ref{fig:a35}(d)] sources according to Eq.~\eqref{eq:1}. Figure~\ref{fig:3}(f) shows that the bulging-out effect results in locally high Joule heating (high $\dot{s}_\mathrm{gen}$) near the bubble tip and locally high heat inflow (high $q_\mathrm{in}$). Figures~\ref{fig:a3}(a)--\ref{fig:a3}(f) in the Supplemental Material~\cite{supp} show the variations of vapor pressure, temperature, density, net heat transfer, interfacial temperature drops, and condensation coefficients of the film bubble for different equilibrium sizes $\theta_\mathrm{eqm}$. As Joule heating is greater near the pore center in the axial direction and greatest at the bubble tip in the radial direction [Fig.~\ref{fig:a35}(c) in the Supplemental Material~\cite{supp}], the highest thermal gradient [Fig.~\ref{fig:a35}(d)] occurs at the bubble tip, resulting in a higher heat and mass inflow into the bubble at the tip. On the other hand, near the bubble base, there is condensation together with partial heat outflux [Fig.~\ref{fig:3}(d)]. The remaining heat input from the liquid is conducted to the surface of the silicon nitride membrane. The fact that the liquid temperature at the LV interface is nonuniform means that there are places where heat and mass flow in and out of the LV interface simultaneously. This is captured mathematically [Eq.~\eqref{eq:14}] by the spatial variation of the condensation coefficient [Fig.~\ref{fig:a3}(h) in the Supplemental Material~\cite{supp}].\par

A stable nanobubble with the simultaneous influx and outflux of dissolved gas on the bubble interface was termed dynamic equilibrium or controlled nonequilibrium by Brenner and Lohse~\cite{Brenner2008}. In that paper, the authors constructed a model based on volume conservation and local gas over-saturation near the solid surface to explain the unusual stability of gas nanobubbles from dissolution. Later, Liu et al.~\cite{liu2017dynamic} performed continuum simulations, showing that simultaneous influx and outflux of dissolved hydrogen on the \textit{gas-liquid} interface can also create a dynamic equilibrium, explaining both the stability of electrochemically generated nanobubbles and the corresponding steady-state current drop on the nanoelectrode. On a similar note, we show here that heat and mass fluxes can also reorganize on the \textit{liquid--vapor} interface of a bubble, lending it stability. We define the term, \textit{dynamic thermal equilibrium} for this mode of thermal nanobubble stability.\par

It should be noted here that the solutions for the dynamic equilibrium size $\theta_\mathrm{eqm}$ for different values of $\alpha_\mathrm{e}$ at the same bias voltage are very close [blue and turquoise traces in Fig.~\ref{fig:3}(h)]. However, as shown in Figs.~\ref{fig:a3}(f) and~\ref{fig:a38}(f) in the Supplemental Material~\cite{supp}, small values of $\alpha_\mathrm{e}$ cause negative and unrealistic values of $\alpha_\mathrm{c}$ in order to satisfy the heat flow from the liquid. Hence, only for high values of evaporation coefficients a stable LV interface can exist. Furthermore, compared to $\alpha_\mathrm{e}=0.5$, when $\alpha_\mathrm{e}=0.9$, we find that $\alpha_\mathrm{c}$ is less non-uniform and varies over a shorter range of 0.82-0.98 on the bubble surface (Fig.~\ref{fig:a3}f and h).\par

Here, we should note that a similar dynamic thermal equilibrium can be caused by the self-organization of other interfacial parameters. For example, $\alpha_\mathrm{c}$ can remain uniform on the LV interface, while $\alpha_\mathrm{e}$ self-organizes to conduct the non-uniform heat fluxes coming from the liquid. To test this hypothesis, we solved for the bubble size under the condition of $\alpha_\mathrm{c}=0.9$ and obtained a spatially varying $\alpha_\mathrm{e}$ ranging from 0.51 to 0.94 (Fig.~\ref{fig:a32}c). When uniform $\alpha_\mathrm{c}=0.9$ is applied on the LV interface, the equilibrium size of $\theta_\text{eqm}=117\degree$ is obtained at 8.04 V (Fig.~\ref{fig:a32}a). On the other hand, when uniform $\alpha_\mathrm{e}=0.9$ is applied on the LV interface, equilibrium size of $\theta_\mathrm{eqm}=117\degree$ is obtained at 8.02 V (Fig.~\ref{fig:2}d). Also, the vapor temperature is 415.91 K when $\alpha_\mathrm{e}=0.9$ (Fig.~\ref{fig:a3}b) and 416.33 K (Fig.~\ref{fig:a32}b) when $\alpha_\mathrm{c}=0.9$. This indicates that we converge to a similar dynamic thermal equilibrium bubble size and temperature from both assumptions. It should be noted that apart from accommodation coefficient variations, variations in vapor molecule velocity distributions on the bubble surface can also accommodate the non-uniform heat fluxes creating a dynamic thermal equilibrium. Our hypothesis of self-organization of interfacial parameters creating a dynamic equilibrium suggests the existence of dissipative structures~\cite{nicolis1977self, kondepudi2008introduction}. The spontaneous evolution of far-from-equilibrium systems to organized states is commonly observed in biology and chemistry. In the case of nanoscale bubbles, the exact molecular mechanism may be very complex, and large-scale nonequilibrium molecular dynamics (NEMD) simulations~\cite{lavino2021surface} might prove useful in resolving the whole picture.\par 

Despite the large heat and mass transport at the bubble interface, dissipation at the bubble interface is negligible compared with the dissipation due to Joule heating and heat conduction in the liquid [Fig.~\ref{fig:a7} in the Supplemental Material~\cite{supp}]. Thus, although the interfacial flux intensifies as the voltage increases to maintain dynamic thermal equilibrium, the increase in dissipation at the bubble interface is much smaller than the decrease in dissipation due to the suppression of Joule heat generation by the expansion of the bubble. Hence, as the voltage is increased, the increasing nonuniform temperature distribution at the bubble interface is accommodated through a dynamic equilibrium mechanism, which locally stabilizes the bubble interface. At the same time, the expansion of the bubble reduces dissipation throughout the system, resulting in thermodynamically stable film boiling.

\subsection{Unsteady states}\label{secIIIC}
\subsubsection{Nucleate boiling}\label{secIIICi}
\begin{figure*}[!t]
%\begin{p}
%\centering
\includegraphics[width=0.67\textwidth,keepaspectratio,angle=0]{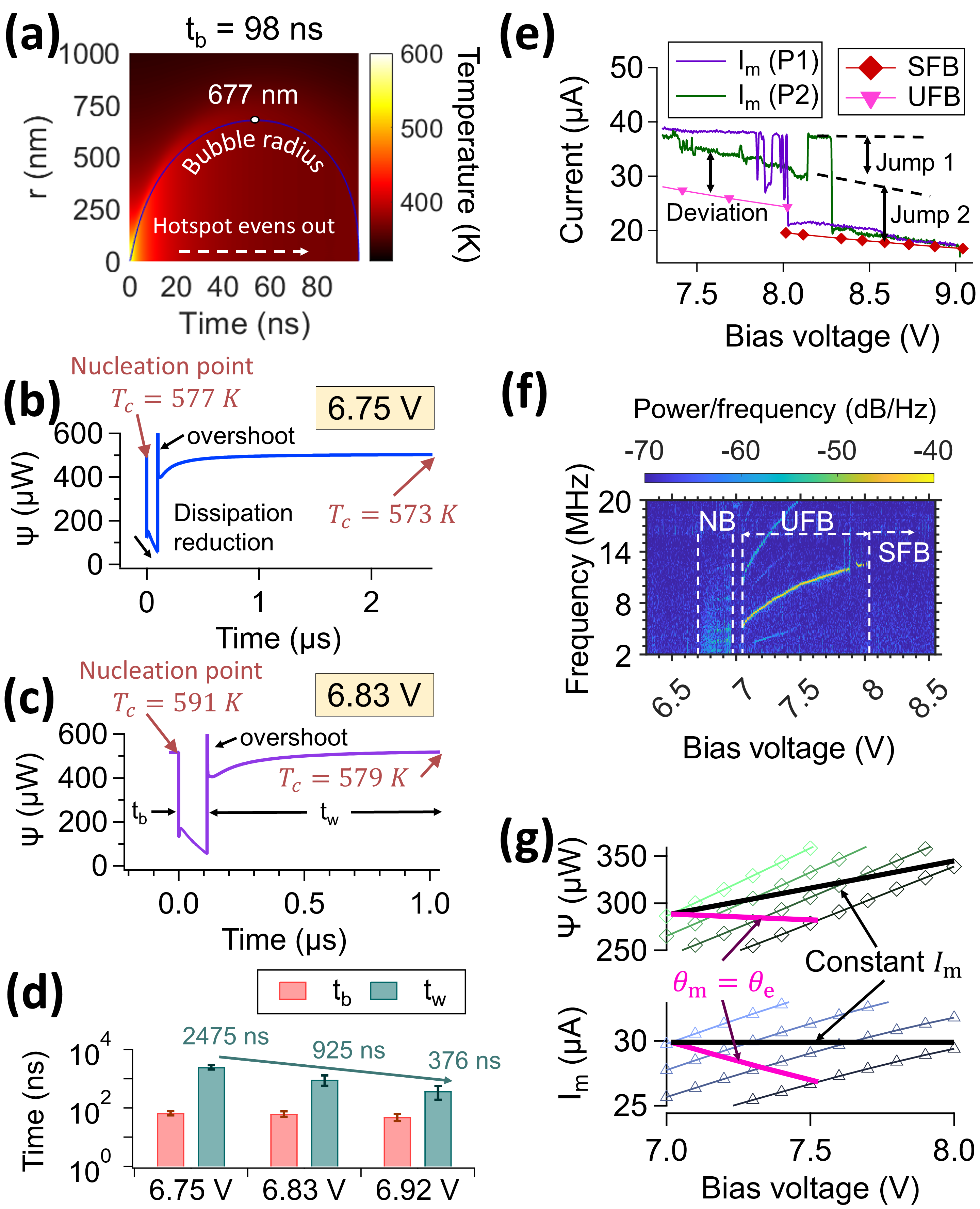}
\caption{\label{fig:4} (a)--(d) Nucleate bubble dynamics. (a) Simulated radial temperature distribution variation during the 98~ns bubble growth--collapse cycle at 6.75~V. (b) and (c) Calculated dissipation functions according to simulations during  periodic nucleate boiling at 6.75~V and 6.83~V, respectively. (d) Variations of blockage durations and waiting time with voltage as seen in experiments. The error bars denote the standard deviations. (e)--(g) Unstable film bubble characteristics. (e) Mean current variation with bias voltage near the second bifurcation point. (f) Current spectrogram of  ramp voltage pulse, P1. (g) Variations of simulated nanopore mean current ($-\triangle-$) and system dissipation ($-\diamond-$) with bias voltage for $\theta=90\degree$, $94\degree$, $98\degree$, and $102\degree$ bubbles in a thermal steady state. Darker shades of blue and green correspond to larger  film bubbles. The pink lines shows the variations of current and dissipation when the bubble size also satisfies mechanical equilibrium. The black line indicates the rise in dissipation when the mean bubble is out of mechanical equilibrium and only in a thermal steady state, albeit satisfying constant mean current.} 
%\end{p}
\end{figure*}

When the linearly increasing voltage reaches 6.7125~V at 14.75~ms, the temperature at the pore center reaches 562~K, and the first homogeneous bubble blockage signal (67~ns in Fig.~\ref{fig:a21} in the Supplemental Material~\cite{supp}) is seen, which marks the onset of nucleate boiling. During this regime, the homogeneous bubble nucleating at the pore center undergoes inertio-thermal bubble growth~\cite{Prosperetti2017, paul2020single}. We simulated the spherical bubble growth--collapse cycle through a one-dimensional moving boundary model~\cite{Robinson2004, paul2021analysis} discussed in our earlier works. A spherical bubble seed is inserted at the pore center, and the hotspot steady-state temperature distribution along the pore axis at 6.75~V is used as the initial radial liquid temperature distribution around the bubble surface. The vapor temperature and density inside the bubble are assumed to follow the saturation line, and the bubble center is fixed at the pore center. Although only radial motion is considered,  good agreement is found in the estimated bubble lifetime of 98~ns compared with the experimental blockage duration of 67~ns. Approximation of radial growth dynamics in the inertio-thermal regime~\cite{sullivan2022inertio} has proven quite useful in simplifying cavitation bubble dynamics~\cite{zhong2020model} and plasmonic bubble dynamics~\cite{zhang2021dynamics}. According to our model~\cite{paul2021analysis}, the bubble radius $R$ reaches a maximum value of 677~nm before shrinking back to the liquid phase [Fig.~\ref{fig:4}(a)]. As the bubble radius was larger than the pore radius for almost the entire lifetime, we assume  total cutoff of Joule heat generation inside the nanopore. This causes a drastic reduction in dissipation post-nucleation [Figs.~\ref{fig:4}(b) and~\ref{fig:4}(c)]. Furthermore, the growth cycle causes  the pre-nucleation hotspot [Fig.~\ref{fig:4}(a)] temperature distribution in the liquid to even-out, leading to a gradual reduction in dissipation during the bubble lifetime [Fig.~\ref{fig:4}(c)]. As the Joule heat generation is disrupted during the bubble lifetime, the sharp temperature distribution in the silicon nitride membrane also relaxes, which was simulated using a one-dimensional conduction model assuming zero heat flux on the pore surface~\cite{paul2020single}. This rough assumption allows us to simplify the complicated conjugate heat transfer problem during nucleate bubble growth and collapse. However, as the dissipation in the silicon nitride membrane has an overall minor contribution to the total system dissipation [Fig.~\ref{fig:a2}c], said assumption should not have a major impact on the period-averaged system dissipation values during nucleate boiling regime (blue marker in Fig.~\ref{fig:2}g).\par

After the bubble growth--collapse cycle, the nanopore Joule heating re-commences, and this is simulated using a transient Joule heating model~\cite{paul2020single} implemented on an axisymmetric finite volume mesh [Fig.~\ref{fig:a22}]. We find that the system dissipation  increases [Figs.~\ref{fig:4}(b) and~\ref{fig:4}(c)] as Joule heating is switched on and the thermal hotspot at the pore center is regenerated after a waiting period [Fig.~\ref{fig:1}(b)], leading  to  subsequent bubble nucleation. From experiments, we found that the blockage duration remained almost the same, while the waiting times between bubble nucleations decreased steadily as the bias voltage was increased [Fig.~\ref{fig:4}(d)]. Accordingly, the period-averaged system dissipation during the nucleate boiling cycle also decreased with increasing bias voltage  (blue trace in Fig.~\ref{fig:a6} in the Supplemental Material~\cite{supp}).

\subsubsection{Transition boiling}\label{secIIICii}

As the pore surface temperatures increase with increasing voltage, nucleation of heterogeneous bubbles becomes likely, and these can coalesce to form  a thin vapor film blanketing the pore surface, leading to a decrease in the baseline current. At 6.83~V,  we find intermittent film bubbles [the FB zone in Fig.~\ref{fig:1}(c)] separating periodic bubble and nucleate bubble blockage signals, which signifies transition boiling. We simulate the film bubble shape and temperature using the Joule heating model as described in Sec.~S2 in the Supplemental Material~\cite{supp}. However, as the film bubble is thinner than the bulged-out bubble in stable film boiling (SFB), we neglect the SV interfacial temperature drop and apply the boundary condition $T_\mathrm{SV}=T_\mathrm{v}$ instead of Eq.~\eqref{eq:12}. Accordingly, a steady-state current of \SI{31}{\micro\ampere} is obtained for the dynamic equilibrium bubble size [Fig.~\ref{fig:3}(c)], which slightly underestimates the experimental nanopore FB current at 6.83~V [Fig.~\ref{fig:1}(c)]. 
% The variation of vapor pressure, temperature, density, net heat transport from liquid to solid, interfacial temperature drops, and condensation coefficients on LV interface are shown in Fig.~\ref{fig:a38} (SI~\cite{supp}).
As the average bubble height satisfies $h_\mathrm{avg}>\lambda$ during unstable film boiling (UFB) [Fig.~\ref{fig:3}(h)], we estimate the temperature drop in the vapor bulk, $\Delta T_\mathrm{bulk}$, using a one-dimensional thermomass model that considers the diffusio-ballistic (non-Fourier) heat transport~\cite{Jou2005, Joshi1993, Dong2012, Sellitto2012, Guo2015}. Details of the simulation results for unstable film bubbles are given in Sec.~S4 in the Supplemental Material~\cite{supp}. However, when the vapor film has $h_\mathrm{ave}=1\lambda$, $\Delta T_\mathrm{bulk}$ increases as the bubble expands with increasing voltage [Fig.~\ref{fig:a38}(e) in the Supplemental Material~\cite{supp}], revealing the limitations of this model. In other words, if the thin vapor film stays in the range $h_\mathrm{ave}<\lambda$, it will not reach mechanical equilibrium and is prone to collapse. This was actually observed in the experiments [Fig.~\ref{fig:1}(c)], where the FB disappears after a few tens of microseconds, allowing  nucleate boiling to resume. We estimated the average system dissipation during intermittent film boiling. Details are given in Sec.~S5 of the Supplemental Material~\cite{supp}. The average system dissipation (green diamond markers in Fig.~\ref{fig:a6}) was obtained by a weighted average of the nucleate boiling dissipation [Figs.~\ref{fig:4}(b) and~\ref{fig:4}(c)] and the unstable film bubble dissipation [Figs.~\ref{fig:a7}(c) and~\ref{fig:a7}(d)], based on their relative probabilities of occurrence $P_\mathrm{NB}$ and $P_\mathrm{FB}$ as read from the experimental current signals.

% (green diamond markers in Fig.~\ref{fig:a6} in the Supplemental Material~\cite{supp}), by weighted averaging of the nucleate boiling dissipation (Fig.~\ref{fig:4}b and c) and the unstable film bubble (UFB) dissipation (Fig.~\ref{fig:a7}c and d) based on their relative probability of occurrence ($P_\mathrm{NB}$ and $P_\mathrm{FB}$) as read from the experimental current signals. 
As can be seen in Fig.~\ref{fig:3}(g), following the first bifurcation, the dissipation starts to decrease with the onset of nucleate boiling, and further decreases during intermittent film boiling (IFB), which  leads overall to a greater reduction in dissipation, albeit one that is not fully stable [Fig.~\ref{fig:1}(d)]. 

In the boiling curve [Fig.~\ref{fig:1}(e)], we find that the Joule heat dissipation rises from 7.1~V to 8~V before the second bifurcation event. The mean current $I_\mathrm{m}$ [the purple trace in Fig.~\ref{fig:4}(e)] flattens in this voltage range, indicating minor bubble expansion. On the other hand, the current spectrogram [Fig.~\ref{fig:4}(f)] reveals the appearance of a distinct frequency band that rises from 5~MHz to 13~MHz in this voltage range. However, as shown in Fig.~\ref{fig:4}(e), in the second ramp voltage pulse (P2), unlike the first ramp voltage pulse (P1), the mean current $I_\mathrm{m}$ decreases monotonically with voltage up to 8.1~V. The spectrogram also shows no frequency band up to 8.1~V at P2 (Fig.~\ref{fig:a5} in the Supplemental Material~\cite{supp}). Therefore, we can infer that the mean bubble position is out of equilibrium (less bulged position), where the bubble is in pinned volumetric self-oscillation~\cite{jenkins2013self, paul2022boiling, Nguyen2018, Li2017a}. In comparison, during nucleate boiling (NB) there is a power spectrum distributed over a wide frequency range, originating from non-sinusoidal current fluctuations comprising unequal blockage durations and waiting times. Accordingly, we can discern the nucleate boiling regime comprising of transient bubble nucleations from the frequency signature as well.\par
Although estimating entropy generation during UFB is quite cumbersome~\cite{otsubo2022estimating}, if we assume that the mean bubble position is determined from dynamic thermal equilibrium while holding $I_\mathrm{m}$ constant, the system dissipation increases with bias voltage [the black line in Fig.~\ref{fig:4}(g)]. This rise in dissipation and Joule heating due to incomplete film bubble expansion constitutes the second rising arm of the ``M''-shaped boiling curve [Figs.~\ref{fig:1}(e) and~\ref{fig:3}(g)]. On the other hand, assuming that the mean bubble position is determined from mechanical equilibrium, the system dissipation decreases with increasing bias voltage [the pink line in Fig.~\ref{fig:4}(g)].

For the P2 pulse, the current rises sharply at 8.1~V (jump 1), from which a small frequency band appears and continues until 8.3~V (Fig.~\ref{fig:a5}), and the current drops sharply, marking the second bifurcation event. The mean current agrees well with the stable film boiling (SFB) simulation results [the red trace in Fig.~\ref{fig:4}(e)], and the oscillation spectrum disappears for both P1 [Fig.~\ref{fig:4}(f)] and P2 (Fig.~\ref{fig:a5}). In the UFB region, the bubble swelling increases $\Delta T_\mathrm{bulk}$, and the heat transport becomes less ballistic, destabilizing the bubble~\cite{seddon2011knudsen}. However, it is unclear why the bubble sometimes starts oscillating at a mean position that is not in equilibrium. Furthermore, for the P2 pulse, a shift was observed between the extrapolated UFB mean current and the SFB current (jump 2). This suggests that a membrane or film bubble with $h_\mathrm{avg}$ in the range from $\lambda$ to $2\lambda$ (at UFB) follows a different heat transport and stabilization mechanism than one with $h_\mathrm{avg}=2\lambda$ (at SFB), causing this quantized behavior. In summary, the second bifurcation follows a stochastic jump process~\cite{van1992stochastic} which can be potentially useful as a switch in nanofluidic computing applications~\cite{noy2023nanofluidic, marr2007micro, karnik2005electrostatic}. However, further studies using large-scale nonequilibrium molecular dynamics  (NEMD) simulations~\cite{lavino2021surface} may help explain the anomalous behavior of the Knudsen membrane bubble.

\section{Conclusions}\label{secIV}
Nanopore Joule heating serves as an excellent platform to detect single-bubble dynamics at nanosecond resolutions. Based on an ``M''-shaped boiling curve for a 199~nm nanopore, superheating, homogeneous nucleate boiling, transition boiling, and stable film boiling regimes are classified. Two bifurcation points, namely, the onset of nucleate boiling and the onset of stable film boiling, are observed, at which the system finds a dissipation reduction mechanism and self-organizes into a more stable boiling structure. During nucleate boiling, the periodicity of bubble nucleation increases \textit{gradually}  with increasing bias voltage, leading to dissipation reduction. At the second bifurcation point, the self-oscillating and entropy-generating film bubble expands \textit{catastrophically} into a bigger film bubble, leading to a stepwise reduction in system dissipation. These bifurcations are explained by theoretical calculations of the variation in dissipation function using a continuum model. The model yields thermal properties of the film bubble as well as nanopore currents that are in good agreement with experimental results. We show that in addition to thermodynamic effects such as dynamic thermal equilibrium and ballistic heat transfer, confinement effects like contact-line pinning must be considered to explain stable film boiling in experiments. Furthermore, our model suggests that a self-organization of thermal accommodation coefficients at the liquid--vapor interface can account for the transport of high heat and mass fluxes through nanoscale bubbles, creating a dynamic thermal equilibrium, which causes nanobubble stability and unique boiling structures in extreme thermal environments. However, a detailed understanding of the molecular dynamics within such Knudsen bubbles is still a challenge, for which further investigation would be required.

%TC:ignore
\begin{acknowledgments}
The authors would like to gratefully acknowledge Juan G Santiago at Stanford University for his valuable comments and discussions on the manuscript. This work was supported by the Japan Society for the Promotion of Science (JSPS) KAKENHI Grant Nos.~20H02081 and 20J22422. Part of this work was conducted at the Advanced Characterization Nanotechnology Platform of the University of Tokyo, which was supported by the Nanotechnology Platform of the Ministry of Education, Culture, Sports, Science, and Technology (MEXT), Japan, Grant No.~JPMXP09A21UT0089.
\end{acknowledgments}
%TC:endignore

\bibliography{refs_EDITED}% Produces the bibliography via BibTeX.

\makeatletter\@input{xx.tex}\makeatother
\end{document}